\documentclass[apj,a4paper]{emulateapj}
\usepackage{natbib}
\usepackage{graphicx} 
\usepackage{amsmath}

\usepackage[normalem]{ulem}

\newcommand{\GS}{{GS95}}

\shorttitle{Diffusion in critically balanced turbulence}
\shortauthors{Laitinen et al.}

\begin{document}

\title{ENERGETIC PARTICLE DIFFUSION IN CRITICALLY BALANCED TURBULENCE}

\author{T. Laitinen, S. Dalla, J. Kelly, and M. Marsh}
\affil{Jeremiah Horrocks Institute, University of Central Lancashire,
  PR1 2HE Preston, UK}

\begin{abstract}
  Observations and modelling suggest that the fluctuations in
  magnetised plasmas exhibit scale-dependent anisotropy, with more
  energy in the fluctuations perpendicular to the mean magnetic field
  than in the parallel fluctuations and the anisotropy increasing at
  smaller scales. The scale-dependence of the anisotropy has not been
  studied in full-orbit simulations of particle transport in turbulent
  plasmas so far. In this paper, we construct a model of critically
  balanced turbulence, as suggested by \cite{GoSr1995}, and calculate
  energetic particle spatial diffusion coefficients using full-orbit
  simulations. The model uses an enveloped turbulence approach, where
  each 2-dimensional wave mode with wavenumber $k_\perp$ is packed
  into envelopes of length $L$ following the critical balance
  condition, $L\propto k_\perp^{-2/3}$, with the wave mode parameters
  changing between envelopes. Using full-orbit particle simulations,
  we find that both the parallel and perpendicular diffusion
  coefficients increase by a factor 2, compared to previous models with
  scale-independent anisotropy.
\end{abstract}


\section{Introduction}\label{sec:introduction}

The origin of Solar Energetic Particles (SEPs) is one of the unsolved
problems in heliospheric physics. Both flares and coronal mass
ejections are capable of accelerating particles, and studies analysing
SEP events give differing conclusions on the main particle
accelerators \cite[see,
e.g.][]{Cane2010,Gopalswamy2012,Aschwanden2012}. The interpretation of
SEP events is complicated due to the particle propagation effects
caused by the turbulent interplanetary magnetic field. SEP events are
observed at a large range of latitudes and longitudes
\citep[e.g.,][]{DallaEa2003Annales,Liu2011,Dresing2012}, suggesting
possible cross-field transport of the SEPs. Thus, in order to
understand the SEP acceleration in the solar eruptions, we must first
understand the nature of SEP tranport in the heliosphere.

Cosmic ray research is typically based on two theoretical
approaches: the diffusion-convection equation \citep{Parker1965}, and
the quasilinear approach (QLT) \citep{Jokipii1966}.  The original
description of the cross-field transport of a charged particle as
propagation in random-walking magnetic field lines \citep{Jokipii1966}
has been improved to take into account the interplay between parallel
and perpendicular propagation effects
\citep{Matthaeus2003,Shalchi2010a}. The spread of the particles
observed in the SEP events, however, remains difficult to explain, as
they require larger ratios of the perpendicular-to-parallel diffusion
coefficients than what can be expected from the theoretical approaches
\citep{Dresing2012}.

Recently, particle transport research has benefited from numerical
simulations utilising particle full-orbit simulations, which have the
advantage of needing no a priori assumptions for the particle
propagation. \citet{Beresnyak2011} and \citet{Wisniewski2012}
  have recently studied particle propagation using MHD-simulated
  turbulence. However, as MHD simulations are limited to small range
  of scales, most full-orbit particle
simulations describe the fluctuating fields as a superposition of
Fourier modes on a constant magnetic field \citep[e.g.][]{GiaJok1999,
  Qin2002, Qin2002_apjl, Zimbardo2006, RuffoloEa2008}.

Observations and modelling suggest that the fluctuations in magnetised
plasmas are anisotropic, with more energy in the fluctuations
perpendicular to the mean magnetic field than in the parallel
fluctuations \citep[e.g.][]{Shebalin1983,Bieber1996}. The anisotropy
is also scale-dependent, with structures at smaller scales more
anisotropic than the larger scales. The scale-dependence of the
anisotropy was predicted by \citet[][hereafter \GS]{GoSr1995}, who
noted that as the turbulence amplitudes increase at small scales, they
introduce variation along the mean field direction. This is due to the
fact that when interacting, the wave packets propagate through each
other, along the mean magnetic field. When the changes in the
interacting wave packets reach nonlinear amplitudes, both of the
interacting waves change, and for this reason also the outcome of the
interaction changes. This takes place on nonlinear timescales
$\tau_{NL}$, when the wave packet's front has propagated distance
$\propto V_A\; \tau_{NL}$, where $V_A$ is the Alfv\'en velocity. Using
the nonlinear timescale from Kolmogorov scaling, \GS\ obtained a
critical balance, $k_\parallel\propto k_\perp^{2/3}$ between the
parallel and perpendicular wavenumbers. The critical balance between
structure scales has been found in MHD turbulence simulations
\citep[e.g.,][]{Cho2002}, and observed in the solar wind turbulence
\citep[e.g.,][]{Horbury2008,Podesta2009}. In order to understand the
propagation of particles in heliospheric plasmas, the scale-dependence
of the turbulence anisotropy should thus be taken into account.

In this work, we study the particle propagation in the critically
balanced turbulence of \GS. To model the turbulence, we use the
enveloping approach of \cite{LaEa2012}. The enveloping approach
improves the earlier models that use infinite, linear plane waves by
breaking the phase coherence of the waves in the direction of the mean
magnetic field. Such a loss of coherence is expected in the plasma in
the heliosphere, as the waves evolve in a nonlinear manner \citep[see,
e.g.,][for a review]{TuMarsch1995}. \cite{LaEa2012} studied the effect
of structures on the tranport and used a fixed envelope length of the
order of turbulence correlation length. In the present paper, we study
the particle propagation in turbulence with scale-dependent
anisotropy, and the enveloping is done separately for each wave mode,
with the envelope length following the critical balance scaling,
$L\propto k_\perp^{-2/3}$, where $L$ is the envelope length and
$k_\perp$ the wavenumber of the enveloped Fourier mode. The Fourier
modes are chosen to be 2D modes, with their wave vector perpendicular
to the mean magnetic field. Thus, the variation of the magnetic field
along the mean field is purely due to the enveloping of the modes. We
use particle full-orbit simulations in the modelled turbulence, and
calculate the diffusion coefficients for energetic particles,
comparing them to results given in previous work.  In Section~2 we
describe the critically balanced turbulence model and the method used
for obtaining the energetic particle diffusion coefficients. In
Section~3, we study the scale-dependence of the turbulence in the
developed model, and report changes in the energetic particle
diffusion coefficients, compared to the traditional composite
model. We discuss the results in Section~4, and draw our conclusions
in Section~5.

\section{Model}\label{sec:model}

\subsection{Turbulence Model}

In our model, the magnetic field consists of a uniform and constant
background field, $B_0$, overlaid with a turbulent field,
$\delta\mathbf{B}(x,y,z)$, with the total magnetic field given as
$$
\mathbf{B}=\mathbf{B}_0 \hat e_z + \delta\mathbf{B}(x,y,z), 
$$
For the background field we use $B_0=5$~nT, consistent with
the magnetic field magnitude at 1~AU.

\begin{figure}
%   \plotone{gs_packs}
   \plotone{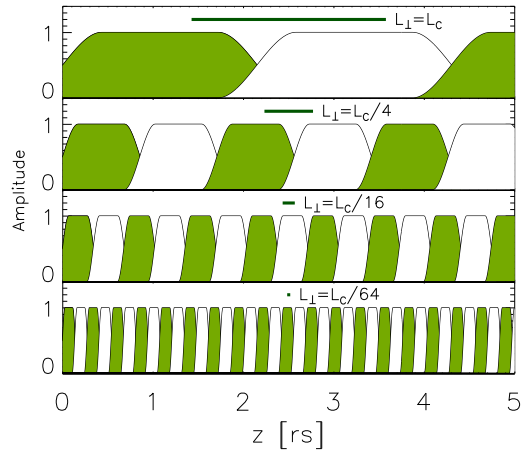}
   \caption{The green and white packets represent the envelopes along
     the mean field direction, for four different perpendicular scales,
     shown by the thick horizontal lines. The steepness parameter is
     $\sigma=10$ in this figure. \label{fig:gspacks} }
 \end{figure}

We model the scale-dependent turbulence by using the turbulence
envelope approach, introduced by \cite{LaEa2012}. In the approach, the
turbulence is enveloped into packets along the mean field direction
(see Fig.~\ref{fig:gspacks}). From an envelope to the next, the
parameters of the wave modes (random phase, wave vector
direction) change. Thus, a particle propagating from an envelope to another
interacts with a coherent Fourier mode only within one envelope, and
the linear coherence is broken when it enters a different envelope.

The scale-dependence of the turbulence in our model is achieved
through the selection of the envelope lengths, which in the study of
\cite{LaEa2012} were constant.
Here the enveloping is done separately for each wave mode, with the
wave scale $\lambda_{\perp n}=2\pi/k_{\perp n}$, where $k_{\perp n}$ is the wavenumber of
the 2D mode. The envelope length, $L_n$, follows the critical balance
scaling, $L_n=L_C^{1/3} \lambda_{\perp n}^{2/3}$, where $L_c$ is the scale
where parallel and perpendicular scales are in balance. As a result,
for wave scales $\lambda_{\perp n}<L_C$ the perpendicular wave scales
decrease faster than the parallel envelope scales, resulting in
scale-dependent anisotropy. This process is depicted in
Fig.~\ref{fig:gspacks}, where, we show the wave scale (thick
horizontal line) compared to the envelope length scales, for four
different wave scales.

The amplitude of mode $n$ for envelope $i$ is modulated by function
$$ \mathfrak{A}_{n,i}(z)= \frac{1}{2}\left[F\left\{\frac{z-z_i}{2S_n}\right\}-F\left\{\frac{z-L_{n}-z_i}{2S_n}\right\}\right]$$ 
where $z_i$ marks the beginning of the envelope and $z$ is the
distance along the direction of the constant magnetic field. The
profile of the envelope is given by function
\begin{equation}
  F\{z\}=
  \begin{cases}
    -1 & z < -1 \\ 
    \frac{3}{2}z-\frac{1}{2}z^3 & -1\leq z\leq 1 \\
    1 & z>1
  \end{cases}
\end{equation}
This approximates the often-used tanh profile, which enables analysing
differentiable profiles that can have steep edges, approaching a
step-function shape. The polynomial description was chosen for numerical
efficiency. The steepness of the envelope profile is given by parameter
$S_n=L_n/\sigma$, and $\sigma$ is is a dimensionless
parameter relating the envelope length to the rise profile length.

In the modelling, we consider that a wave mode with a given wave
vector direction and random phase gives its energy to another mode
with the same modulus of $k_n$, without losses, thus keeping the sum
of the amplitudes of the modes constant,
$\sum_i\mathfrak{A}_{n,i}(z)=1$. This is achieved by selecting
$z_{i+1}=z_i+L_n$.

 \begin{figure}
%   \plotone{gs_packs_profiles}
   \plotone{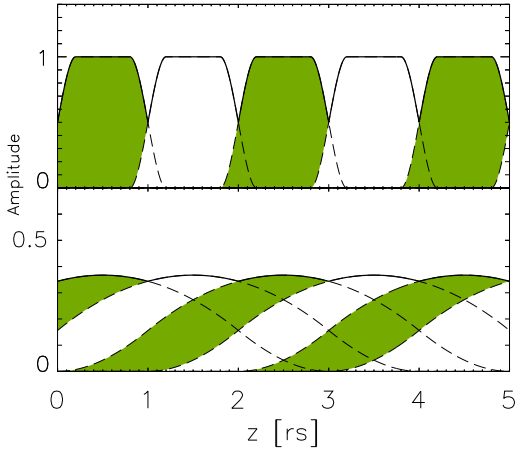}
  \caption{The green and white packets represent the
              envelopes along the mean field direction, for
              $\sigma=10$ (top panel) and $\sigma=1$ (bottom
              panel). The envelope length $L_{n,i}=1$ for these
              profiles.\label{fig:gspackprofiles} }
 \end{figure}

 The steepness of the envelope is related to the transfer rate of
 energy from a mode to another, due to the critical balancing of the
 turbulence. The rate of change depends on the relative wavenumber
 directions of the interacting modes \citep[e.g.][]{LuoMelrose2006},
 which results in some interactions transferring the energy faster
 than others. In our model, however, we describe phenomenologically
 only the result of the critically balanced 3-wave interactions, thus
 accurate description of the rise profile is not possible on this
 level. Instead, we consider in this study two different profiles, a
 gradual change with $\sigma=1$ and a rapid change, with
 $\sigma=10$. As shown in Fig.~\ref{fig:gspackprofiles}, for
 $\sigma=10$ the energy is transferred from one mode to another in an
 intermittent manner, whereas for $\sigma=1$ the interaction is more
 gradual, between multiple modes. The distance along the mean field
 direction over which one mode dominates over the others, remains the
 same, the envelope length, $L_n$, for both values of $\sigma$.

The turbulence field is given as a sum of $M_{n}$ envelopes and $N$
Fourier modes, as
$$ \delta
\mathbf{B}(x,y,z)=\sum_{n=1}^{N}\sum_{i=1}^{M_{n}}\mathfrak{A}_{n,i}(z)\;A(k_n)
\hat{\mathbf{\xi}}_{n,i}\mathrm{e}^{i(k_{\perp n} z'_{n,i}+\beta_{n,i})}.$$        
In this study the Fourier modes in the envelopes are 2D modes, with
the wave vector $\mathbf{k}_{\perp n}$ perpendicular to the mean magnetic
field. Thus the variation on parallel scales in the model are due to
the enveloping only. The polarisation vector $\hat\xi_{n,i}$ lies in
the $xy$-plane and is perpendicular to both the mean magnetic field
and $\mathbf{k}_n$, in order to satisfy $\nabla \cdot
{\mathbf{B}}=0$. Thus
$$\hat\xi_{n,i}=i \hat{\mathbf{y}}'_n$$
where the coordinate system $\mathbf{r}'$ is obtained from $\mathbf{r}$
with a rotation matrix \citep{GiaJok1999}.
The wave propagation directions in the $xy$-plane, as well as the
random phases $\beta_{n,i}$, are chosen from a uniform random
distribution.

The fluctuation amplitude $A(k_n)$ is given by a power law spectrum
\begin{equation}
 A^2(k_n) =  B_1^2 \frac{G_n}{\sum_{n=1}^{N} G_n},
 \;\; G(k_n) = \frac{\Delta V_n}{1+(k_n L_c )^\gamma}, \label{eq:spectrum}
\end{equation}
where $B_1^2$ is the variance of the magnetic field, $L_c$ the
spectrum's turnover scale, $\gamma$ the spectral index, and $\Delta
V_n$ specifies the volume element in $k$-space that the discrete mode
$k_{\perp n}$ represents. For the turnover scale we use $L_c=2.15\, r_\odot$,
where $r_\odot$ is the solar radius. For 2D turbulence, the spectral
index is 8/3 and the factor $\Delta V_n=2\pi k_n\Delta k_n$. In the
simulations we use logarithmically spaced wavemodes with the
wavenumber running from $2\pi/1 \mathrm{AU}$ to $2\pi/10^{-4}
\mathrm{AU}$.

To compare the turbulent fields and their effects on SEP propagation,
we scale the fluctuation amplitudes so that the average energy density
in the fluctuating field is independent of the enveloping
parameters. The scaling factor is obtained through numerical
integration.

\subsection{Energetic Particle Simulations}\label{sec:partsim}

We study particle propagation in the modelled turbulent magnetic
fields by integrating the fully relativistic equation of motion of
energetic protons using the simulation code by \citet{Dalla2005}.  The
code uses the Bulirsh-Stoer method \citep{NumRecp}, with adaptive
timestepping to control the accuracy by limiting the error between
consecutive steps to a given tolerance. We simulate 2048 particles in
10 field realisations, with different wave mode random phases and wave
vector directions in each realisation, thus giving a total of 20480
particles. Each realisation has N=128 Fourier modes, whereas the
number of envelopes depends on the energy of the particle. The
particles are simulated for $\sim100$ parallel diffusion times.
The particle diffusion coefficient is obtained with
\begin{equation}
\kappa_{\zeta\zeta}=\frac{\left<\Delta\zeta^2\right>}{2t},
\,\,\,\,\,\zeta=x,y,z,\label{eq:diffcoeff}
\end{equation}
\citep[e.g.][]{GiaJok1999}, with
$\kappa_{\parallel}$  equal to $\kappa_{zz}$, whereas
$\kappa_{\perp}$ is obtained as the mean of $\kappa_{xx}$ and
$\kappa_{yy}$. For
  further details of the simulation scheme, and verification against
  the results of \citet{GiaJok1999}, see
  \citet{LaEa2012}.

\section{Results}

\subsection{Scale-dependence of the turbulence}

\begin{figure}
%   \plotone{gs_corr}
   \plotone{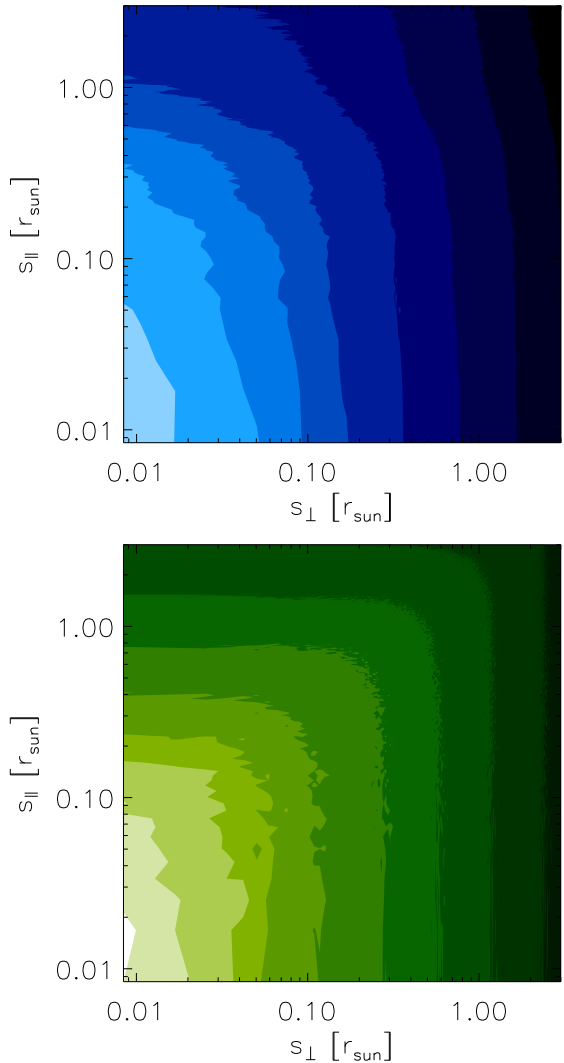}
  \caption{Two-point correlation
              functions for the composite model (top panel),
              and for the \GS\ turbulence model (bottom
                panel).\label{fig:corrcontours} }
 \end{figure}

The goal of this work is to study particle motion in a turbulent field
that has a scale-dependent anisotropy that corresponds to the
critically balanced scaling suggested by \GS. As the first step, we
compare the composite turbulence model of \cite{GiaJok1999} to the one
presented in this study. We do this by calculating the two-point
correlation function,
\begin{equation}
  C_{xx}(\mathbf{s})=\left<\delta
    B_x(\mathbf{r})\delta
    B_x(\mathbf{r}+\mathbf{s})\right>
\end{equation}
where $\mathbf{s}$ is the scale vector the correlation is calculated
for, and the angle brackets denote ensemble average. We calculate the
correlation function as a function of $s_\perp$ and $s_\parallel$,
that is, across and along the mean field direction. The correlation
function is calculated from the dissipation scales to the turnover
scale $L_c$ of the spectrum, and presented in
Fig.~\ref{fig:corrcontours}, where the contours are selected to
represent correlations at equidistant scales, $s$, rather than values
of the correlation function, $C_{xx}$. The correlation contours for
the \GS\ model are ellipses (appearing rectangular in the log-log
presentation of Fig.~\ref{fig:corrcontours}), with aspect ratio
increasing at smaller scales. A similar trend cannot be seen in the
composite model.

 \begin{figure}
%   \plotone{scaledep_anisotr}
   \plotone{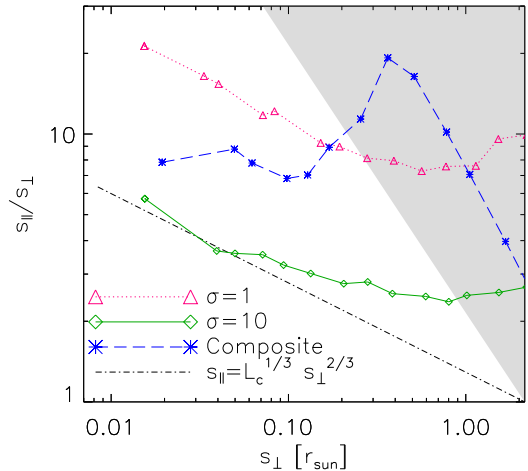}
   \caption{ The
     scale-dependence of the two models with the axis ratio
     $s_\parallel/s_\perp$. In the grey area,
     $L_c^{1/3}s_\perp^{2/3}$ exceeds the turnover scale
     $L_c$.\label{fig:scaledep} }
 \end{figure}

To quantify the scale-dependence of the anisotropy, we calculate the
axis ratio of the contours. This is done by finding values of the
parallel and perpendicular scales, $s_\parallel$ and $s_\perp$, that
satisfy $C_{xx}(s_{\perp},0)=C_{xx}(0,s_{\parallel})$. We present the
axis ratio $s_{\parallel}/s_{\perp}$ as a function of $s_\perp$, for
the turbulence models, in Fig.~\ref{fig:scaledep}. The dash-dotted
straight line represents the \GS\ scaling, the solid and the dotted
curves (green and red in the online version) represent the
phenomenological model of this work, with two different values of the
steepness parameter $\sigma$, and the dashed curve (blue in the online
version) represents the composite model of \citet{GiaJok1999}. As can
be seen, the axis ratio in the model presented in this work is clearly
scale-dependent, and follows the trend of the \GS\ scaling. There is a
significant dependence of the level of anisotropy on the steepness
parameter, with the steep profile for the energy change between the
wavemodes, represented by $\sigma=10$, producing values closer to
isotropy when the axis scales approach the isotropy scale $L_c$.

At large scales, the phenomenological model deviates from the \GS\
model. This is due to the different form of the turbulence spectrum
used in the derivation of the critical balance by \GS, and the one
used in this study. The \GS\ scaling is obtained by using a continuous
power law spectrum, $k_\perp^{5/3}$, representing the Kolmogorov
turbulence inertial range spectrum, while in the heliosphere the
inertial range is limited in extent to only a few orders of magnitude,
as the spectrum flattens at large scales into energy-containing range. This
results in flattening of the correlation function at scales of order
$L_c$ and larger. Thus, when $\lambda_\perp < L_\parallel \lesssim
L_c$, with $L_\parallel=L_c^{1/3} \lambda_\perp^{2/3}$, the parallel
scale $L_\parallel$ reaches the flattening of the correlation function
before the perpendicular scale $\lambda_\perp$, and the ratio
$L_\parallel/\lambda_\perp$ remains anisotropic. This range is shown
by gray shading in Fig.~\ref{fig:scaledep}. As the spectrum defined by
Eq.~(\ref{eq:diffcoeff}) rolls gradually to the inertial range, it is
expected to see the deviation from the \GS\ form to start already at
somewhat smaller scales. Solar wind turbulence observations show a
somewhat smaller, but still anisotropic axis ratio of 1.5 for scales
of $\sim r_\odot$ for slow solar wind \citep{Dasso2005}.

Thus, we conclude that our model with $\sigma=10$ displays the required
scale-dependence of anisotropy, with the large scale anisotropy being
consistent with the solar wind observations of \cite{Dasso2005}. The
gradual energy transfer model, $\sigma=1$, produces higher level of
anisotropy, exceeding the \cite{Dasso2005} result.

\subsection{Turbulence spectrum}

 \begin{figure}
%    \plotone{spectrastrips}
    \plotone{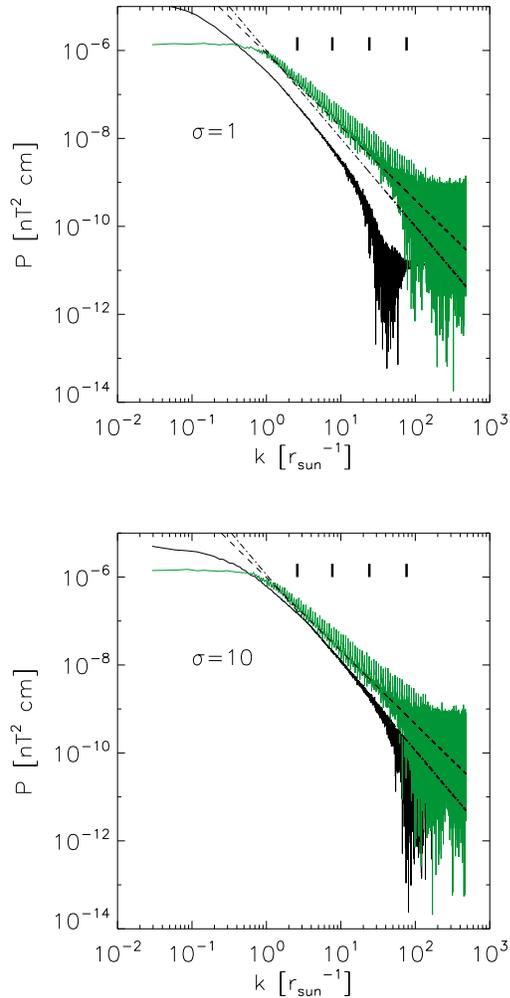}
    \caption{\label{fig:spectra} Parallel (black curve) and
      perpendicular (grey curve, green in the online version) power
      spectrum of the \GS\ turbulence model with $\sigma=1$ (top panel)
      and $\sigma=10$ (bottom panel).The dash-dotted and dashed
        curves depict the theoretical spectra given by
        Eqs.~\ref{eq:gs_par} and \ref{eq:gs_perp}, respectively. The
        vertical bars at the top of the panels depict the inverse of
        the larmor radii of protons of energy 100, 10, 1 and 0.1~MeV,
        respectively, from left to right.}
  \end{figure}

In the previous section, we described the scale-dependent behaviour of
the two-point correlation function for different turbulence
models. When studying turbulence, it is often more common to
present a power spectrum, which quantifies the energy deposited in the
turbulence at different scales. We present such spectra in
Fig.~\ref{fig:spectra}. The
spectra are obtained from one-dimensional two-point correlation
functions, with parallel spectrum defined as 
$$P_{xx}(k_\parallel)=\int C_{xx}(0,s_\parallel)\, e^{i k_\parallel s_\parallel}
ds_\parallel,$$
and the $P_{xx}(k_\perp)$ similarly from
$C_{xx}(s_\perp,0)$. The one-dimensional spectrum can be
  compared to the three-dimensional spectral form given by \GS\ as
\begin{equation*}
P(\mathbf{k})=C\frac{V_{A}^{2}}{k_{\perp}^{10/3}L^{1/3}}\, g\left(\frac{k_{\parallel}L^{1/3}}{k_{\perp}^{2/3}}\right),
  \label{eq:gs_spectrum}
\end{equation*}
where $C$ is a dimensionless constant, $L$ the isotropic excitation
scale, and $g(x)$ a function that vanishes at large $x$. From this
form the one-dimensional spectrum can be obtained using the Taylor
hypothesis \citep{Taylor1938},
\begin{equation*}
P(f)=\int d^{3}\mathbf{k}\, P(\mathbf{k})\delta(2\pi f-\mathbf{k}\cdot\mathbf{V}).  \label{eq:Taylor}
\end{equation*}
It is straigthforward to show that the one-dimensional spectrum along
the field line is proportional to $k^{-2}$ and the perpendicular
spectrum to $k^{-5/3}$. Furthermore, we find that when using the
simpler form suggested by \cite{Cho2002}, $g(x)=\exp(-x)$, the spectra
are given in form
\begin{eqnarray}
P_{xx}(k_\parallel)&=&\frac{3}{2}C\left(\frac{1}{k_{\parallel}L}\right)^{2}\,
V_{A}^{2}\, L \label{eq:gs_par}\\
 P_{xx}(k_\perp)&=&C\left(\frac{1}{k_{\parallel}L}\right)^{5/3}\, \label{eq:gs_perp}
V_{A}^{2}\, L.
\end{eqnarray}
We plot these spectra in Fig.~\ref{fig:spectra}, using $L_c$ for the
isotropic excitation scale $L$, and fitting the perpendicular
spectrum, Eq.~(\ref{eq:gs_perp}), to the perpendicular spectrum of the
simulated turbulence. It should be noted that the parallel spectrum
represents a continuum instead of being composed of discrete modes, as
it results from the enveloping rather than begin generated with a sum
of parallel Fourier modes.

As seen in the figures, the envelope shape affects the parallel
spectrum shape. For the gradual envelope profile, with $\sigma=1$
(Fig.~\ref{fig:spectra}, top panel), the parallel spectrum has
spectral index of -2.5, steeper and an order of magnitude
  below the spectrum given by Eq.~(\ref{eq:gs_par}).  For the steep
profile, with $\sigma=10$ (Fig.~\ref{fig:spectra}, bottom panel),
the parallel spectrum has a spectral index of $-2$, consistent
  with Eq.~(\ref{eq:gs_par}), with the relative power in the parallel
  and perpendicular spectra consistent with Eqs.~(\ref{eq:gs_par})
  and~(\ref{eq:gs_perp}). The perpendicular spectrum is not affected
by the enveloping, with the spectral index remaining at the Kolmogorov
value of -5/3, the input of the model for both values of $\sigma$.
The spectral index of -2 for the parallel direction and -5/3 for the
perpendicular direction were recently observed in the solar wind by
\cite{Horbury2008} \cite{Podesta2009}. The ratio $P_\perp/P_\parallel$
in our model is 2 at $k=10\,\, r_\odot^{-1}$ and 5.7 at $100\,\,
r_\odot^{-1}$, which is similar to the results of \cite{Horbury2008}
and \cite{Podesta2009}, who find values of ~1.5--5 for the range
$k=10-1000 \,\,r_\odot^{-1}$.

\subsection{Energetic Particle Propagation}\label{sec:sep_propagation}

  \begin{figure}
%    \plotone{kappa_gs_ene_steeps}
    \plotone{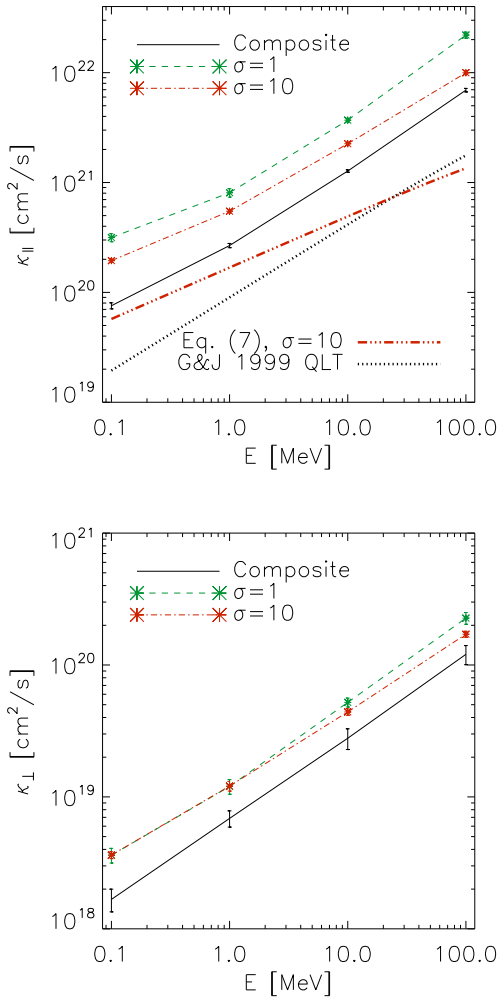}
    \caption{\label{fig:kappa}Parallel (top panel) and perpendicular
      (bottom panel) energetic particle diffusion coefficients in
      composite turbulence (solid curve), and in the \GS\ turbulence
      (dashed and dash-dotted curves, green and red in the online
      version), as a function of proton energy, for turbulence
      amplitude $B_1/B_0=1$. In the top panel, the dashed
        curve depicts the QLT diffusion coefficient for the composite
        model, as given by \citet{GiaJok1999}, and the tripledot-dash
        line Eq.~\ref{eq:qlt} for the
        $\sigma=10$ case.}
 \end{figure}

 In order to understand how the scale-dependence of turbulence
 anisotropy affects the energetic particle propagation, we have
simulated energetic protons at energies 0.1, 1, 10 and 100
   MeV, with the particles Larmor radius scale, $k_L=r_L^{-1}$ shown
   as vertical bars in Fig.~\ref{fig:spectra}. To study the collective
   behaviour of the particles, we have calculated the particle
 diffusion coefficients along and across the mean magnetic field, as
 described in Section~\ref{sec:partsim}, shown in
   Fig.~\ref{fig:kappa}, with statistical error limits. For
 comparison, we have also followed particles in the composite model
 turbulence of \citet{GiaJok1999}. The parallel diffusion coefficients
 for the \GS\ model and the composite model are presented in the top
 panel of Fig.~\ref{fig:kappa}. In the figure, the solid curve
 represents the composite model, whereas the dashed and dash-dotted
 curves (green and red in the online version) represent the \GS\
 model, for two values of the steepness parameter. As can be seen, the
 parallel diffusion coefficient is larger for the \GS\ model, and
 depends strongly on the steepness parameter of the scale-dependent
 enveloping. This can be understood on the basis of lower energy in
 parallel fluctuations to scatter the particles in the $\sigma=1$ case
 compared to the $\sigma=10$ case. The dependence of the parallel
 diffusion coefficient on particle energy in the \GS\ model also
 differs from the composite model, with the diffusion coefficient
 decreasing slower at decreasing energies. This can be understood as
 the effect of the steeper spectrum, which results in smaller relative
 energy in the small-scale fluctuations that scatter the lower-energy
 particles.
 
 We present also theoretical estimations of the parallel
   diffusion coefficient in Fig.~\ref{fig:kappa}. The dotted line
   represents the QLT $\kappa_\parallel,$ as given by
   \cite{GiaJok1999} presented in their work for comparison with the
   composite and isotropic turbulence models. Their estimation is
   based on an equation that is valid only for parallel Alfv\'en waves
   (the slab waves). It should be noted that the in a composite
   turbulence the 2D component does not have a significant
   contribution to parallel scattering \citep[e.g.,][]{Bieber1994}. If we
   assume that only the slab component contributes to the parallel
   scattering, the 80\%:20\% mix of 2D and slab modes implies a factor
   5 rise of the dotted curve in the upper panel of
   Fig.~\ref{fig:kappa}, which fits well to the simulated composite
   model results.

 Comparison of the theoretical parallel diffusion coefficient
   with our results for the \GS\ turbulence is not straightforward, as
   the QLT for magnetostatic turbulence results in infinite diffusion
   coefficient for parallel spectrum $\propto k_\parallel^{-2}$ and
   steeper. This resonance gap problem, due to lack of scattering at
   low particle pitch angle cosines, $\mu=v_\parallel/v$, has been
   studied in the scientific literature, and several mechanisms for
   its closure have been suggested
   \citep[e.g.,][]{SchlickeiserAchatz1993,Bieber1994,NgReames1995,Vainio2000}.
   As shown by, e.g.,~\citet{Vainio2000}, if scattering at low pitch
   angles after the closure of the resonance gap is weak, the
   diffusion coefficient is dominated by this low rate of scattering,
   and thus by the closure mechanism. If, however, the resonance gap
   is efficiently closed, the pitch angle diffusion coefficient can be
   estimated by
 \begin{equation}
   \kappa_\parallel\propto v r_L \frac{B_0^2}{r_L^{-1} P(r_L^{-1})} h(\mu_g) \label{eq:qlt}
 \end{equation}
 where $r_L$ is the particle's Larmor radius, and $h(\mu_g)$ is a
 dimensionless function that depends on the spectral index of the
 parallel turbulence and the pitch angle $\mu_g$ below which the
 resonance gap closure mechanisms overcome the QLT result. This form
 is valid only for a power law spectrum of the turbulence, and thus
 the higher energies in this study are expected to have larger
 diffusion coefficients, compared to the estimation.  In addition, it
 should be noted that the above considerations are valid for slab
 turbulence. As our model contains a wealth of oblique wave modes,
 implied by the correlation function, the QLT would give a wealth of
 harmonic resonances in addition to the resonance at inverse Larmor
 radius.

We present the theoretical parallel diffusion coefficient, as
  given by Eq.~(\ref{eq:qlt}), in Fig.~\ref{fig:kappa}, with a
  tripledot-dash line, for our model with steepness parameter
  $\sigma=10$. The power spectrum in Eq.~(\ref{eq:qlt}) is obtained as
  a power law fit to the parallel spectrum
  (Fig.~\ref{fig:spectra}). The exact scaling, given by $h(\mu_g)$, is
  not known, thus we can only compare the trends of the diffusion
  coefficients. As can be seen, the QLT trend tracks the \GS\ model
  only at low energies. This suggests that the resonance gap effects
  must be taken into account when estimating the mean free path at
  steep spectra.

 The perpendicular diffusion coefficients are shown in the bottom
 panel of Fig.~\ref{fig:kappa}. The perpendicular diffusion is stronger in
 the \GS\ turbulence, by a factor of 2, as compared to the composite
 model. The dependence on the enveloping shape is small except for the
 highest energies used in the study. Thus, the perpendicular and
 parallel diffusion coefficients behave differently with respect to
 the exact description of the critically balanced turbulence.

\section{Discussion and Conclusions}

In this work, we study how energetic particles propagate in critically
balanced turbulence. We build turbulence with scale-dependent
anisotropy resulting from critical balance by using envelopes of
length that follows the required scaling between the perpendicular
wavenumber and the parallel variation scale. Analysis of the 2-point
correlation function and the spectra shows that for the envelope
steepness parameter $\sigma=10$ our model is consistent with the \GS\
critical balance scaling and the observations of solar wind turbulence
\citep{Dasso2005,Horbury2008,Podesta2009}. The gradual shape,
$\sigma=1$ (see Fig.~\ref{fig:gspackprofiles}) results in more
pronounced anisotropy. The case of $\sigma=10$, however, can be
considered more realistic, as the nonlinear interactions between the
waves depend on their relative phases \citep[e.g.,][]{LuoMelrose2006},
resulting in an uneven evolution of the wave packets. The structure of
rapid change after an extended period of unchanged turbulence can also
be viewed as intermittent turbulence. We note, however, that in our
model the probability distribution of magnetic field increments,
$\Delta
B_x(s_\perp,s_\parallel)=B_x(r_\perp+s_\perp,r_\parallel+s_\parallel)-B_x(r_\perp,r_\parallel)$
does not significantly deviate from Gaussian, unlike for the observed
and modelled turbulence \citep[e.g.,][]{Greco2009}.

The scale-dependence of the turbulence results in a notable difference
in the power spectra along and across the mean magnetic field, as
shown by Fig.~\ref{fig:spectra}. Thus, it is expected that the
energetic particle transport differs from the composite model of
\cite{GiaJok1999}, where the spectral indices are both equal to
5/3. Our simulations show that the effect for the parallel diffusion
coefficient is significant, with a 1.5--2.5 -fold increase, depending
on the particle energy, for $\sigma=10$ (top panel of
Fig.~\ref{fig:kappa}). An increase of the diffusion coefficient with a
steepening parallel spectrum is consistent with the standard
quasilinear theory \citep[e.g.,][]{DungSchlickeiser1990}. Our
  analysis shows, however, that the simple estimate based on QLT, for
  the parallel wave modes (Eq.~(\ref{eq:qlt})) fails to describe the
  parallel diffusion coefficient at higher energies.

The perpendicular propagation is also affected by the scale-dependence
of the turbulence, with the perpendicular diffusion coefficient larger
by a factor 2 compared to the one obtained with composite model. This
can be understood on the basis of the work of \cite{Matthaeus2003}. In
their formulation, the perpendicular diffusion coefficient depends on
the parallel diffusion coefficient. Thus, a different form for the
parallel component spectrum would influence the cross-field transport
of the energetic particles.  \cite{Shalchi2010b} used the
further-developed model by \cite{Shalchi2010a} to calculate the
perpendicular diffusion coefficient in a \GS-type spectrum. Their
$\kappa_\perp/\kappa_\parallel$ decreases $\sim 5$-fold compared to
the composite model, and deviates from our result, where the ratio
does not change for $\sigma=10$. However, in their work they did not
estimate the parallel diffusion coefficient consistently using the
\GS-spectrum, but used standard quasilinear theory \citep{Jokipii1966}
with Kolmogorov spectrum, and a model based on Galactic cosmic ray
observations. Thus, their results are not comparable to
$\kappa_\perp/\kappa_\parallel$ in our model. The insensitivity of
$\kappa_\perp$ to the parallel diffusion coefficient in their
model is in line with our results of insensitivity of $\kappa_\perp$
to the changes in the turbulence caused by the change of the envelope
shape. This insensitivity can be implied also from the theoretical
work by \cite{Matthaeus2003} (their Eq.~(7)), where the rapidly
decreasing power spectrum may diminish the link between the parallel
and perpendicular diffusion.

Determining the correct scattering parameters in the interplanetary
space is very important for the analysis of SEP events.  Studies have
shown that the scattering along the interplanetary magnetic field can
significantly affect the accuracy of the SEP event onset analysis at 1
AU \citep{LintunenVainio2004,Saiz2005,LaEa2010}.  Recently,
\cite{GiaJok2012} showed that varying the diffusion coefficients and
the ratio of the perpendicular and parallel diffusion coefficient in
the simulations can result in a wide variety of intensity profiles at
1 AU for an impulsive SEP event.  Therefore the two-fold increase in
the parallel and perpendicular transport coefficients reported in this
paper will significantly influence the characteristics of SEP events.

\acknowledgments

We acknowledge support from the UK Science and Technology Facilities
Council (STFC) (grant ST/J001341/1) and from the European
Commission FP7 Project COMESEP (263252). JK acknowledges STFC support
via Ph.D. studentship.

\bibliographystyle{apj}

\begin{thebibliography}{41}
\expandafter\ifx\csname natexlab\endcsname\relax\def\natexlab#1{#1}\fi

\bibitem[{{Aschwanden}(2012)}]{Aschwanden2012}
{Aschwanden}, M.~J. 2012, \ssr, 171, 3

\bibitem[{{Beresnyak} {et~al.}(2011){Beresnyak}, {Yan}, \&
  {Lazarian}}]{Beresnyak2011}
{Beresnyak}, A., {Yan}, H., \& {Lazarian}, A. 2011, \apj, 728, 60

\bibitem[{{Bieber} {et~al.}(1994){Bieber}, {Matthaeus}, {Smith}, {Wanner},
  {Kallenrode}, \& {Wibberenz}}]{Bieber1994}
{Bieber}, J.~W., {Matthaeus}, W.~H., {Smith}, C.~W., {Wanner}, W.,
  {Kallenrode}, M.-B., \& {Wibberenz}, G. 1994, \apj, 420, 294

\bibitem[{{Bieber} {et~al.}(1996){Bieber}, {Wanner}, \&
  {Matthaeus}}]{Bieber1996}
{Bieber}, J.~W., {Wanner}, W., \& {Matthaeus}, W.~H. 1996, \jgr, 101, 2511

\bibitem[{{Cane} {et~al.}(2010){Cane}, {Richardson}, \& {von
  Rosenvinge}}]{Cane2010}
{Cane}, H.~V., {Richardson}, I.~G., \& {von Rosenvinge}, T.~T. 2010, JGR (Space
  Physics), 115, 8101

\bibitem[{{Cho} {et~al.}(2002){Cho}, {Lazarian}, \& {Vishniac}}]{Cho2002}
{Cho}, J., {Lazarian}, A., \& {Vishniac}, E.~T. 2002, \apj, 564, 291

\bibitem[{{Dalla} \& {Browning}(2005)}]{Dalla2005}
{Dalla}, S., \& {Browning}, P.~K. 2005, \aap, 436, 1103

\bibitem[{{Dalla} {et~al.}(2003){Dalla}, {Balogh}, {Krucker}, {Posner},
  {M{\"u}ller-Mellin}, {Anglin}, {Hofer}, {Marsden}, {Sanderson}, {Heber},
  {Zhang}, \& {McKibben}}]{DallaEa2003Annales}
{Dalla}, S., {et~al.} 2003, Annales Geophysicae, 21, 1367

\bibitem[{{Dasso} {et~al.}(2005){Dasso}, {Milano}, {Matthaeus}, \&
  {Smith}}]{Dasso2005}
{Dasso}, S., {Milano}, L.~J., {Matthaeus}, W.~H., \& {Smith}, C.~W. 2005,
  \apjl, 635, L181

\bibitem[{{Dresing} {et~al.}(2012){Dresing}, {G{\'o}mez-Herrero}, {Klassen},
  {Heber}, {Kartavykh}, \& {Dr{\"o}ge}}]{Dresing2012}
{Dresing}, N., {G{\'o}mez-Herrero}, R., {Klassen}, A., {Heber}, B.,
  {Kartavykh}, Y., \& {Dr{\"o}ge}, W. 2012, \solphys, 281, 281

\bibitem[{{Dung} \& {Schlickeiser}(1990)}]{DungSchlickeiser1990}
{Dung}, R., \& {Schlickeiser}, R. 1990, \aap, 237, 504

\bibitem[{{Giacalone} \& {Jokipii}(1999)}]{GiaJok1999}
{Giacalone}, J., \& {Jokipii}, J.~R. 1999, \apj, 520, 204

\bibitem[{{Giacalone} \& {Jokipii}(2012)}]{GiaJok2012}
---. 2012, \apjl, 751, L33

\bibitem[{{Goldreich} \& {Sridhar}(1995)}]{GoSr1995}
{Goldreich}, P., \& {Sridhar}, S. 1995, \apj, 438, 763

\bibitem[{{Gopalswamy} {et~al.}(2012){Gopalswamy}, {Xie}, {Yashiro}, {Akiyama},
  {M{\"a}kel{\"a}}, \& {Usoskin}}]{Gopalswamy2012}
{Gopalswamy}, N., {Xie}, H., {Yashiro}, S., {Akiyama}, S., {M{\"a}kel{\"a}},
  P., \& {Usoskin}, I.~G. 2012, \ssr, 171, 23

\bibitem[{{Greco} {et~al.}(2009){Greco}, {Matthaeus}, {Servidio}, {Chuychai},
  \& {Dmitruk}}]{Greco2009}
{Greco}, A., {Matthaeus}, W.~H., {Servidio}, S., {Chuychai}, P., \& {Dmitruk},
  P. 2009, \apjl, 691, L111

\bibitem[{{Horbury} {et~al.}(2008){Horbury}, {Forman}, \&
  {Oughton}}]{Horbury2008}
{Horbury}, T.~S., {Forman}, M., \& {Oughton}, S. 2008, Physical Review Letters,
  101, 175005

\bibitem[{{Jokipii}(1966)}]{Jokipii1966}
{Jokipii}, J.~R. 1966, \apj, 146, 480

\bibitem[{{Laitinen} {et~al.}(2012){Laitinen}, {Dalla}, \& {Kelly}}]{LaEa2012}
{Laitinen}, T., {Dalla}, S., \& {Kelly}, J. 2012, \apj, 749, 103

\bibitem[{{Laitinen} {et~al.}(2010){Laitinen}, {Huttunen-Heikinmaa}, \&
  {Valtonen}}]{LaEa2010}
{Laitinen}, T., {Huttunen-Heikinmaa}, K., \& {Valtonen}, E. 2010, Twelfth
  International Solar Wind Conference, 1216, 249

\bibitem[{{Lintunen} \& {Vainio}(2004)}]{LintunenVainio2004}
{Lintunen}, J., \& {Vainio}, R. 2004, \aap, 420, 343

\bibitem[{{Liu} {et~al.}(2011){Liu}, {Luhmann}, {Bale}, \& {Lin}}]{Liu2011}
{Liu}, Y., {Luhmann}, J.~G., {Bale}, S.~D., \& {Lin}, R.~P. 2011, \apj, 734, 84

\bibitem[{{Luo} \& {Melrose}(2006)}]{LuoMelrose2006}
{Luo}, Q., \& {Melrose}, D. 2006, \mnras, 368, 1151

\bibitem[{{Matthaeus} {et~al.}(2003){Matthaeus}, {Qin}, {Bieber}, \&
  {Zank}}]{Matthaeus2003}
{Matthaeus}, W.~H., {Qin}, G., {Bieber}, J.~W., \& {Zank}, G.~P. 2003, \apjl,
  590, L53

\bibitem[{{Ng} \& {Reames}(1995)}]{NgReames1995}
{Ng}, C.~K., \& {Reames}, D.~V. 1995, \apj, 453, 890

\bibitem[{{Parker}(1965)}]{Parker1965}
{Parker}, E.~N. 1965, \planss, 13, 9

\bibitem[{{Podesta}(2009)}]{Podesta2009}
{Podesta}, J.~J. 2009, \apj, 698, 986

\bibitem[{Press {et~al.}(1993)Press, Teukolsky, Vetteling, \&
  Flannery}]{NumRecp}
Press, W.~H., Teukolsky, S.~A., Vetteling, W.~T., \& Flannery, B.~P. 1993,
  Numerical Recipes in Fortran: The art of Scientific Computing, 2nd edn. (New
  York, NY, USA: Cambridge University Press)

\bibitem[{{Qin}(2002)}]{Qin2002}
{Qin}, G. 2002, PhD thesis, University of Delaware

\bibitem[{{Qin} {et~al.}(2002){Qin}, {Matthaeus}, \& {Bieber}}]{Qin2002_apjl}
{Qin}, G., {Matthaeus}, W.~H., \& {Bieber}, J.~W. 2002, \apjl, 578, L117

\bibitem[{{Ruffolo} {et~al.}(2008){Ruffolo}, {Chuychai}, {Wongpan}, {Minnie},
  {Bieber}, \& {Matthaeus}}]{RuffoloEa2008}
{Ruffolo}, D., {Chuychai}, P., {Wongpan}, P., {Minnie}, J., {Bieber}, J.~W., \&
  {Matthaeus}, W.~H. 2008, \apj, 686, 1231

\bibitem[{{S{\'a}iz} {et~al.}(2005){S{\'a}iz}, {Evenson}, {Ruffolo}, \&
  {Bieber}}]{Saiz2005}
{S{\'a}iz}, A., {Evenson}, P., {Ruffolo}, D., \& {Bieber}, J.~W. 2005, \apj,
  626, 1131

\bibitem[{{Schlickeiser} \& {Achatz}(1993)}]{SchlickeiserAchatz1993}
{Schlickeiser}, R., \& {Achatz}, U. 1993, Journal of Plasma Physics, 49, 63

\bibitem[{{Shalchi}(2010)}]{Shalchi2010a}
{Shalchi}, A. 2010, \apjl, 720, L127

\bibitem[{{Shalchi} {et~al.}(2010){Shalchi}, {B{\"u}sching}, {Lazarian}, \&
  {Schlickeiser}}]{Shalchi2010b}
{Shalchi}, A., {B{\"u}sching}, I., {Lazarian}, A., \& {Schlickeiser}, R. 2010,
  \apj, 725, 2117

\bibitem[{{Shebalin} {et~al.}(1983){Shebalin}, {Matthaeus}, \&
  {Montgomery}}]{Shebalin1983}
{Shebalin}, J.~V., {Matthaeus}, W.~H., \& {Montgomery}, D. 1983, Journal of
  Plasma Physics, 29, 525

\bibitem[{{Taylor}(1938)}]{Taylor1938}
{Taylor}, G.~I. 1938, Royal Society of London Proceedings Series A, 164, 476

\bibitem[{{Tu} \& {Marsch}(1995)}]{TuMarsch1995}
{Tu}, C.-Y., \& {Marsch}, E. 1995, \ssr, 73, 1

\bibitem[{{Vainio}(2000)}]{Vainio2000}
{Vainio}, R. 2000, \apjs, 131, 519

\bibitem[{{Wisniewski} {et~al.}(2012){Wisniewski}, {Spanier}, \&
  {Kissmann}}]{Wisniewski2012}
{Wisniewski}, M., {Spanier}, F., \& {Kissmann}, R. 2012, \apj, 750, 150

\bibitem[{{Zimbardo} {et~al.}(2006){Zimbardo}, {Pommois}, \&
  {Veltri}}]{Zimbardo2006}
{Zimbardo}, G., {Pommois}, P., \& {Veltri}, P. 2006, \apjl, 639, L91

\end{thebibliography}

\end{document}